\documentclass[aps,pra,twocolumn,showpacs]{revtex4-1}

\usepackage{amsfonts,amssymb,amsmath}
\usepackage{mathtools}
\usepackage[]{graphics,graphicx,epsfig}
\usepackage{amsthm,multirow}
\usepackage{color}
\usepackage{xcolor}

\newcommand{\rmv}[1]{\iffalse #1\fi}

\begin{document}

\title{Quantum illumination with non-Gaussian states: Bounds on the minimum error probability using quantum Fisher information}

\author{Changsuk Noh}
\email[]{cnoh@knu.ac.kr}
\affiliation{Kyungpook National University, Daegu 41566, Korea}

\author{Changhyoup Lee}
\affiliation{Quantum Universe Center, Korea Institute for Advanced Study, Seoul 02455, Korea}

\author{Su-Yong Lee}
\email[]{suyong2@add.re.kr}
\affiliation{ Agency for Defense Development, Daejeon 34186, Korea}

\date{\today}

\begin{abstract}
Quantum illumination employs entangled states to detect a weakly reflective target in a thermal bath. The performance of a given entangled state is evaluated from the minimum error probability in the asymptotic limit, which is compared against the optimal coherent state scheme. We derive an upper bound as well as a lower bound on the asymptotic minimum error probability, as functions of the quantum Fisher information. The upper bound can be  achieved using a repetitive local strategy. This allows us to compare the optimal performance of definite-photon-number entangled states against that of the coherent states under local strategies. 
%For a given photon number, we find that the optimum definite-photon-number entangled state is outperformed by the coherent state with the same signal energy.
When optimized under the constraint of a fixed total energy, we find that a coherent state outperforms the definite-photon-number entangled state with the same signal energy.
\end{abstract}

\maketitle

\section{Introduction}
The aim of quantum illumination (QI) is to detect the presence of a weakly reflective target embedded in a thermal background \cite{Lloyd}. Surprisingly, it was shown that an entangled Gaussian quantum probe, namely the two-mode squeezed vacuum (TMSV) state, enhances the performance of the task and that the enhancement is maximal for low signal energy and large thermal noise \cite{Tan}.
Many theoretical  developments \cite{Shapiro09, Guha, Zhang, Zhuang17a, Zhuang17b, Liu, Fan, Ray, Pirandola19, Karsa1, Karsa2, Yung, Blakely, Jeffers, Lee21, Jo, Lee21b,Jo21} and experiments \cite{Genovese,Zheshen,England,Aguilar,Sussman,Xu} have followed in the subsequent decade, which are well documented in reviews \cite{Pirandola, Shapiro, Sorelli}. An extension to radio-wave frequency has been proposed and proof-of-principle experiments have been demonstrated \cite{Sandbo,Shabir15,Shabir19,Luong,Cai}. 

 The two-mode squeezed vacuum  state has been proven to be optimal in the asymmetric (Neyman-Pearson) case \cite{Palma} and nearly optimal in the symmetric case \cite{Ranjith,Bradshaw}.  The corresponding optimal measurements require a collective measurement over $M$ copies of the input state as $M\rightarrow \infty$ and are thus difficult to implement \cite{Zhuang17a}. For local measurements Sanz et al.~derived an achievable upper bound on the minimum error probability \cite{Sanz} using the concept of quantum Fisher information (QFI) extensively used in quantum estimation theory \cite{Paris, Pirandola, Polino}. The use of QFI makes it easier to go beyond the Gaussian regime and investigate the potential benefits of using non-Gaussian entangled states. 
 
 In this work, we provide a simple derivation of the same upper bound on the minimum error probability as a function of the QFI. The same method also provides a loose lower bound.  Using these, we investigate possible advantages of using definite-photon number entangled states of the form $\sum_{n=0}^{N} a_n |N-n,n\rangle_{SI}$, which we will call $N$-photon entangled ($N$PE) states.  These states have been shown to be useful for quantum estimation purposes \cite{Dorner2009}. Possible advantages of using $N$PE states in QI were first investigated in Ref.~\cite{Lee21}, where an optimized $N$PE state was shown to outperform the coherent state under repetitive local measurement strategies. The optimum $N$PE state, for a given value of $N$, was found by numerically maximizing the QFI. Furthermore, by considering a specific non-optimal scheme, a performance measure (signal-to-noise ratio) was shown to increase with a thermal photon number  for large enough values of the latter. Here, we show that, contrary to the findings in Ref.~\cite{Lee21}, the QFIs for the optimized $N$PE states are slightly lower than those of the coherent states when the signal energy is constrained to be equal.   By performing an exact calculation, we also show that the signal-to-noise ratio decreases monotonically as thermal noise is increased.

 \section{Quantum Illumination}
\begin{figure}[h]
\includegraphics[width=0.3\textwidth]{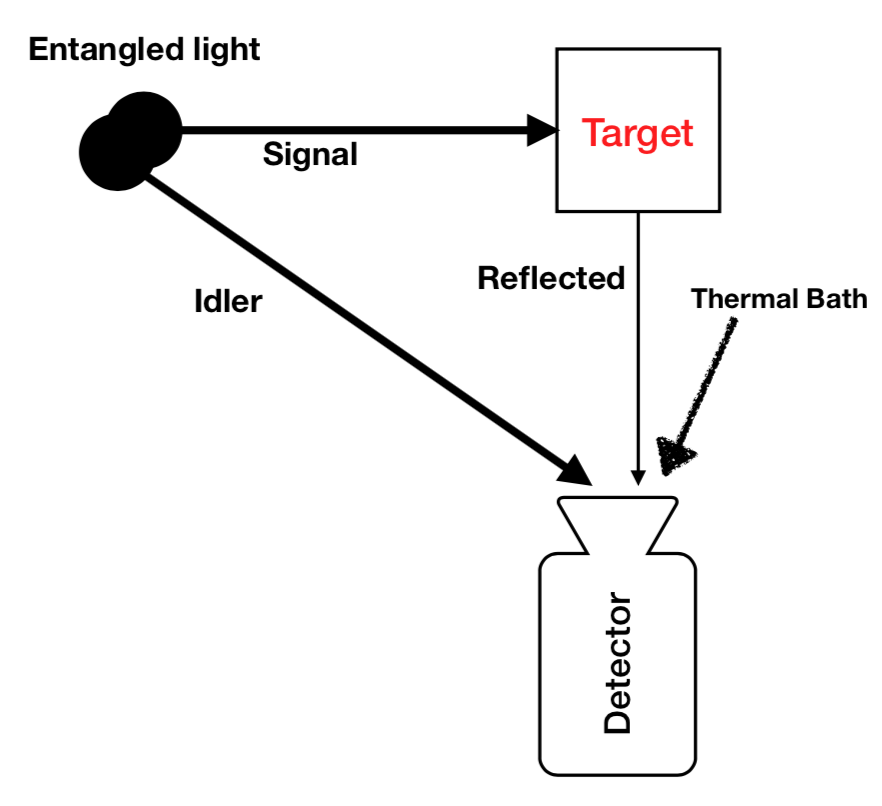}
\caption{Generic quantum illumination scheme. A signal mode of a two-mode entangled light is sent to a weakly reflective target embedded in a thermal bath, while an idler mode remains protected. The reflected signal mode from the target is measured together with the idler mode at a detector. The presence of the target is inferred from the measurement result.} \label{fig:setup_gen}
\end{figure}

Figure \ref{fig:setup_gen} illustrates a general QI protocol. The central goal is to detect the presence of a target that may or may not be there. The target is modeled by a weakly reflective beam splitter with reflectivity $\eta \ll 1$ and the thermal noise is modeled as a separate field entering the unused port of the beam splitter.  Explicitly, the unitary transformation describing the beam splitter can be written as $U_\eta \approx \exp[\eta(a_S^\dagger b - a_S b^\dagger)]$ for small $\eta$, where $a_S$ and $b$ are the annihilation operators for the signal and thermal photons. After receiving the reflected signal, the quantum state is given by $\rho_\eta = {\rm Tr}_{\text{S}} [U_\eta |\psi\rangle\langle \psi| \otimes \rho_{\rm th}U_\eta ^\dagger]$, where $\rho_{\rm th} = \sum_n [n_{\rm th}^n/(1+n_{\rm th})^{n+1}] |n\rangle \langle n|$ is the thermal state with the mean photon number $n_{\rm th}$. $|\psi\rangle $ is an input two-mode entangled state that can be written as $\sum_k \sqrt{p_k} |\psi_k\rangle_S |\phi_k\rangle_I$ in the Schmidt form. $S$ and $I$ denote signal and idler modes, respectively. The goal is then to distinguish between the two density operators $\rho_\eta$ (target present) and $\rho_0$ (target absent). Following the language of hypothesis testing, we label the two cases $H_1$ and $H_0$ respectively.

The task at hand is thus to come up with the best decision strategy to distinguish between the two hypotheses, or in other words to reduce the error probability. There are two types of error. The false-alarm probability $P_{\rm false} = P(1|0)$ is the probability that $H_1$ is chosen when $H_0$ is true, i.e., the object is assumed to be present when it is not. The miss probability $P_{\rm miss} = P(0|1)$ is the probability that $H_0$ is chosen when $H_1$ is true, i.e., the object is assumed to be absent when it is present. The total error probability is then
\begin{align}
P_{\rm err} = \pi_0 P_{\rm false} + \pi_1 P_{\rm miss},
\end{align}
 where $\pi_0$ and $\pi_1$ are the prior probabilities of the object being absent ($H_0$) or present ($H_1$), respectively.  Different optimal strategies exist depending on how the errors are minimized, because the two types of errors cannot be reduced independently. Two types of approaches to hypothesis testing are widely used.
  
 In Bayesian hypothesis testing, one aims to reduce both types of errors simultaneously and minimize the total error probability $P_{\rm err}$. The lowest achievable $P_{\rm err}$ is given be \cite{Helstrom}
 \begin{align}
 P_{\rm err}  =  \frac{1}{2}\left( 1 - || \pi_0 \rho_0 - \pi_1 \rho_1 ||_1\right) ,
 \end{align}
where $\rho_1 \equiv \rho_\eta$ and $||A||_1 = {\rm Tr} \sqrt{A^\dagger A}$  is the trace norm of an operator $A$. 

In  the Neyman-Pearson approach one aims to minimize one of the errors, say $P_{\rm miss}$, while keeping the other, $P_{\rm false}$, below a certain tolerance threshold. This approach is more useful for certain cases, but will not be considered  further in this work. Instead, we concentrate on symmetric Bayesian hypothesis testing, where $\pi_0 = \pi_1 = 1/2$.  
   
Given $M$ copies of the input state, the decision strategies fall into two categories. The first is to perform `non-local' measurements on the collective states, for which the performance measure depends on $ ||  \rho_0^{\otimes M} -  \rho_1^{\otimes M} ||_1$. This quantity is difficult to  evaluate, but fortunately the error probability is bounded by the so-called quantum Chernoff bound as $P_{\rm err} \le Q^M(\rho_0,\rho_1)/2$ where $Q(\rho_0,\rho_1) = {\rm inf}_{0 \le s \le 1} {\rm Tr}(\rho_0^s \rho_1^{1-s})$ \cite{Aud07}. The bound is known to be tight in the asymptotic limit of $M \rightarrow \infty$. Using the fact that $ {\rm Tr}(\rho_0^s \rho_1^{1-s})$ can be  calculated analytically for Gaussian states \cite{Stefano08}, Tan et al.~evaluated the quantum Chernoff bound for the  coherent state: $P_{\rm err}^{\rm coh} \le e^{-M\eta^2 (\sqrt{1+n_{\rm th}}- \sqrt{n_{\rm th}} ) }/2$ \cite{Tan}. For $n_{\rm th} \gg 1$ the right hand side becomes $e^{-M\eta^2 N_S/4n_{\rm th} }/2$. For the TMSV state, they derived a non-tight upper bound at $s = 1/2$, the quantum Bhattacharyya bound, which reduces to $P_{\rm err}^{\rm TMSV}  \le e^{-M\eta^2 N_S/n_{\rm th} }/2$ in the limit of $\eta \ll 1$, $N_S \ll 1$, and $n_{\rm th} \gg 1$. Thus QI with a TMSV state was shown to be able to achieve a 6dB (4 times larger) improvement in the error exponent over the classical (coherent state) illumination. The optimality of the TMSV state, under a joint measurement strategy, has been established recently \cite{Ranjith,Bradshaw}.

The second option is to perform `local' measurements on  individual copies of the received states and form a decision with the $M$ results. Local strategies can be further divided into adaptive and non-adaptive ones, depending on whether measurement results are used to modify another measurement. In this work we only consider a particular types of non-adaptive strategy, where the same measurement is used throughout. This type of measurement strategy is called a repetitive strategy.

\section{Minimum error probability for repetitive local strategies}

Consider a generic local measurement that yields an unbiased estimator of $\eta$. For a given input state, the measurement results follow a certain probability distribution. If the measurements are repeated $M \gg 1$ times, the mean value of the estimator follows a Gaussian distribution according to the central limit theorem. The upshot is that the error probability scales as  \cite{Guha} (see also \cite{Jo} for a more detailed explanation)
\begin{align}
\label{eq3}
P_{\rm err} \sim e^{-M ({\rm SNR})^2/2},
\end{align}
where the signal-to-noise ratio, SNR, is defined as 
\begin{align}
{\rm SNR} = \frac{N_1 - N_0 }{\sigma_1 + \sigma_0}.
\end{align}
Here $N_m$ is the expectation value and $\sigma_m$ is the standard deviation of the estimator assuming hypothesis $H_m$. Hypothesis $H_0$ is chosen if the average value of the estimator after $M$ measurements is below the threshold value $N_{\rm thr} = \frac{\sigma_1 N_0 + \sigma_0 N_1}{\sigma_0 + \sigma_1}$, while $H_1$ is chosen otherwise. Here we use this expression to obtain an upper bound and a lower bound on the error probability as functions of the QFI.

To obtain the bounds we use a well-known inequality between the error probability and the fidelity
\begin{align}
\label{fidel_eqn}
\frac{1-\sqrt{1-\mathcal{F}(\rho_0,\rho_1)^{2M}}}{2} \le P_{\rm err}(\rho_0,\rho_1) \le \frac{\mathcal{F}(\rho_0,\rho_1)^M}{2},
\end{align}
as well as a relation between the fidelity  and the QFI 
\begin{align}
\mathcal{F}(\rho_\theta,\rho_{\theta+d\theta})  = 1-\frac{F_{\rm Q}(\rho_\theta) d\theta^2}{8}.
\end{align}
Here $F_{\rm Q}$ denotes the QFI and the fidelity is defined as $\mathcal{F} (\rho_0,\rho_1) = {\rm Tr} \sqrt{\sqrt{\rho_1}\rho_0\sqrt{\rho_1}}$. 
Because $\eta \ll 1$, we can write 
\begin{align}
\mathcal{F} (\rho_0,\rho_\eta) = 1- \frac{F_{\rm Q}(\rho_0) \eta^2}{8} \approx e^{- \frac{F_{\rm Q}(\rho_0) \eta^2}{8}}.
\end{align}
Plugging this into Eq.~(\ref{fidel_eqn}), we obtain
\begin{align}
\frac{1}{4} e^{-\frac{M F_{\rm Q}(\rho_0) \eta^2}{4}} \lesssim P_{\rm err}(\rho_0,\rho_\eta) \lesssim  \frac{1}{2} e^{-\frac{M F_{\rm Q}(\rho_0) \eta^2}{8}}.
\end{align}
The upper bound coincides with the one derived in Ref.~\cite{Sanz}. There, both $P_{\rm false}$ and $P_{\rm miss}$ were shown to scale as $\exp [-\eta^2 F_{\rm Q} M/8]$ and the corresponding optimal local measurement was shown to be the one that gives the QFI in the estimation scenario. That is, the optimal observable is $\hat{L}/F_{\rm Q}$ where $\hat{L}$ is the symmetric logarithmic derivative of $\rho_\eta$ at $\eta=0$. We note that the bounds in Eq.~(\ref{fidel_eqn}) are not known to be tight, so our derivation does not guarantee that the bound can be achieved.
 
To summarize, we have derived
\begin{align}
\frac{1}{4} e^{-\frac{M F_{\rm Q}(\rho_0) \eta^2}{4}} \lesssim  P_{\rm err} \sim e^{-\frac{M(SNR)^2}{2}} \lesssim \frac{1}{2} e^{-\frac{M F_{\rm Q}(\rho_0) \eta^2}{8}},
\end{align}
for local repetitive strategies when $M\gg1$ and $\eta \ll 1$. Using these inequalities, the QFI can be used to put bounds on the minimum error probability. Even if the measurement is not optimal, one can use the SNR to obtain the error probability via Eq.~(\ref{eq3}). From Ref.~\cite{Sanz}, we know that the upper bound is achievable while the lower bound is generally known to be loose. As an application of this result we study performances of optimal $N$PE states by calculating their QFI, $F_{\rm Q}^{N {\rm PE}}$.
 
\subsection{Quantum illumination with $N$PE states}
\begin{figure}
\includegraphics[width=0.95\columnwidth]{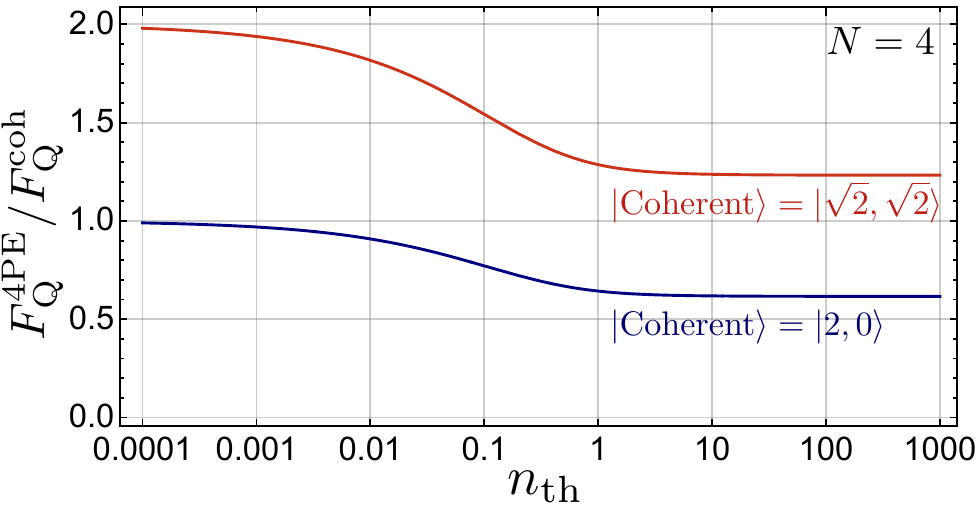} 
\caption{Ratio  between the QFIs, $F_{\rm Q}^{\text{4PE}}/F_{\rm Q}^{\rm coh}$, for an optimized 4PE state and coherent states$|\sqrt{2},\sqrt{2}\rangle$ and $|2,0\rangle$, as functions of the thermal photon number.}
\label{4pes_old}
\end{figure}

In Ref.~\cite{Lee21}, Lee et al.~showed that an optimized $N$PE state has a larger QFI than the coherent state with the same total energy when the energy is evenly distributed between the two modes. That is, when the input coherent state is fixed as $|\sqrt{N},\sqrt{N}\rangle$. For example, with $N=4$, 4PE state has a larger QFI than the coherent state $|\sqrt{2},\sqrt{2}\rangle$ as shown in Fig.~\ref{4pes_old}. This, however, is not the optimal case for the product coherent state with fixed total energy $N$. By concentrating all the energy $N$ on the signal mode, the QFI for the coherent state becomes larger than that of the optimized 4PE state, except at $n_{\rm th}=0$ at which point the QFIs are equal. This is illustrated in Fig.~\ref{4pes_old}.

\begin{figure}
\includegraphics[width=0.92\columnwidth]{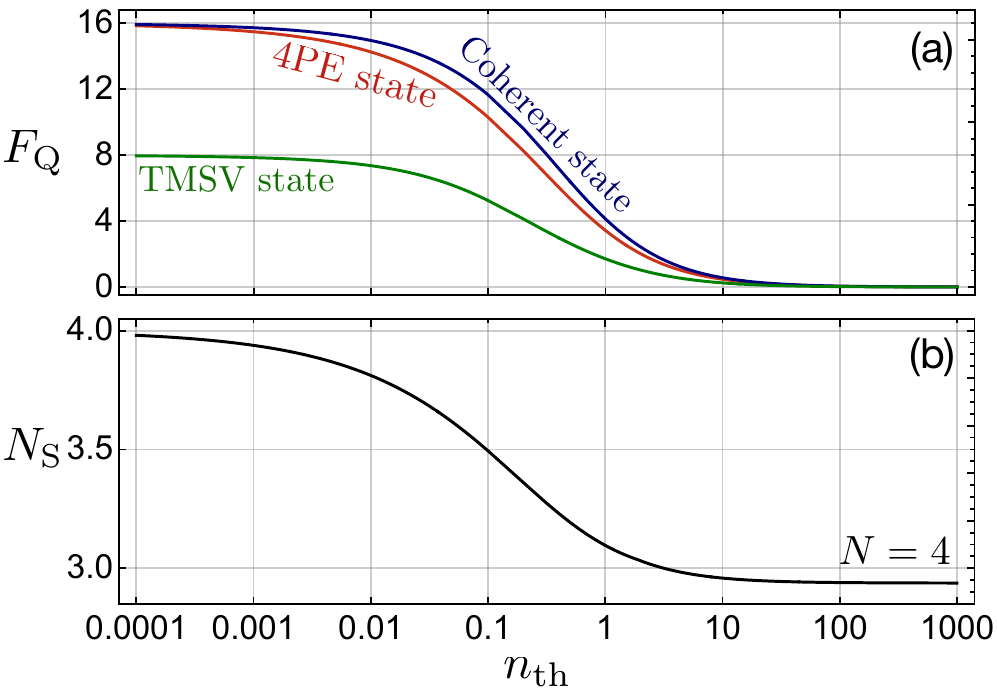}
\caption{Quantum Fisher information as functions of the thermal photon number $n_\text{th}$. (a) $F_{\rm Q}$ for the optimized 4PE state, the coherent state and the TMSV state.  All states have the same signal energy $N_\text{S}$.  (b) The signal energy, i.e., the mean photon number, as a function of $n_{\rm th}$.}
\label{4pes_new}
\end{figure}

Actually, only the signal energy is relevant for separable states since the QFI only depends on it. In particular, the QFI of the separable coherent state $|\alpha,\beta\rangle $ goes as $F_{\rm Q}^{\rm coh} = 4N_S/(1+2n_{\rm th})$ with $N_S = |\alpha|^2$. 
Figure \ref{4pes_new}(a) displays the QFIs for the 4PE state, a single-mode coherent state $|\alpha\rangle$, and the TMSV, for the same values of the signal energy.  To obtain the curve for the 4PE state, we have numerically optimized over the coefficients $\{a_n\}$. The latter can be taken to be real because the phases can be absorbed into the definitions of the idler states without affecting the QFI.  Contrary to the result in Ref.~\cite{Lee21}, where the joint coherent state $|\sqrt{2},\sqrt{2}\rangle$ has been assumed, there is no advantage in using the 4PE states. In fact, the coherent state performs slightly better. The QFI  for the TMSV state is significantly lower than the others and will not be discussed further. In Fig.~\ref{4pes_new}(b), the signal energy $N_S$ is shown as a function of $n_{\rm th}$. It quickly decreases from the maximum value 4 as $n_{\rm th}$ increases from 0 to 1 and then very slowly decreases to a value slightly less than 3. This signifies that in the presence of noise it is better to distribute some of the energy in the $N$PE state to the idler mode, in order to observe correlations with the signal mode.

\begin{figure}
\includegraphics[width=1\columnwidth]{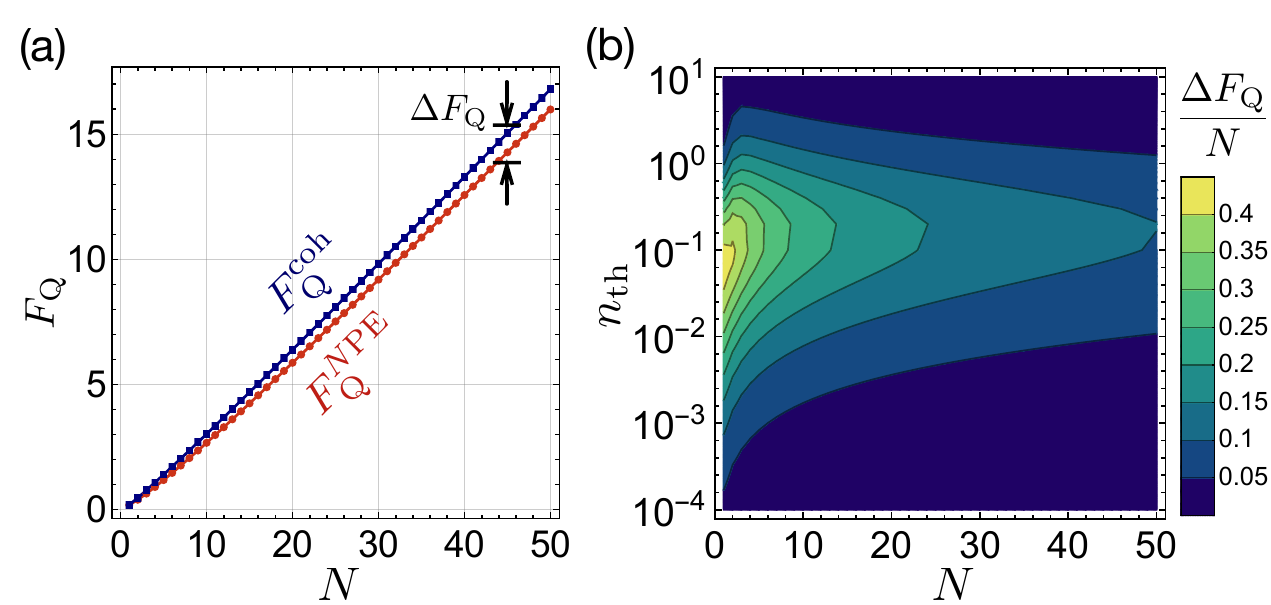}
\caption{\rmv{Differences between $F_{\rm Q}^{\rm coh}$ and $F_{\rm Q}^{N\text{PE}}$.} 
(a) Quantum Fisher information for the coherent state ($F_{\rm Q}^{\rm coh}$) and the optimized 4PE state ($F_{\rm Q}^{N\text{PE}}$) as functions of $N$, for $n_{\rm th} = 5$.
(b) Normalized difference between the QFIs, $\Delta F_\text{Q}/N = (F_\text{Q}^{\rm coh}-F_\text{Q}^{N\text{PE}})/N$ as functions of $N$ and $n_{\rm th}$. }
\label{$N$PES_QFI}
\end{figure}

Figure \ref{$N$PES_QFI}(a) shows how the QFI changes with $N$ for $n_{\rm th} = 5$. It increases almost linearly in $N$ for both states and $F_{\rm Q}^{\rm coh} > F_{\rm Q}^{N {\rm PE}}$  for all values of $N$. The differences between the QFIs, $\Delta F_{\rm Q} = F_{\rm Q}^{\rm coh} - F_{\rm Q}^{N {\rm PE}}$ increases with $N$ and eventually saturates, showing that the difference becomes insignificant for a large value of $N$. Other values of $n_{\rm th}$ produce similar graphs with differing gradients, where the gradient decreases with increasing $n_{\rm th}$.  Figure \ref{$N$PES_QFI}(b) illustrates how $\Delta F_{\rm Q}/N$ changes as a function of $n_{\rm th}$ and $N$, showing that the difference is the most significant around $n_{\rm th} \approx 0.1$ and $N \lesssim 1$.

\begin{figure}
\includegraphics[width=0.9\columnwidth]{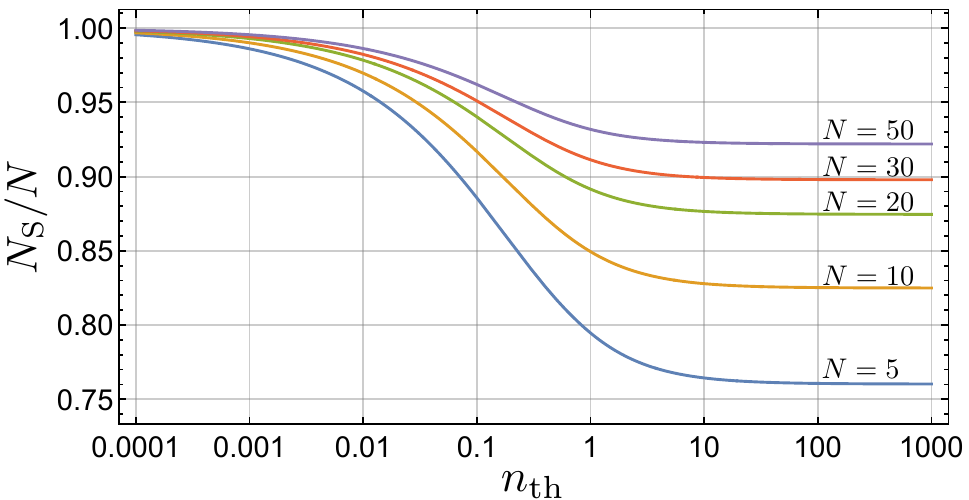}
\caption{Fraction of the total energy carried by the signal mode $N_S/N$ as a function of $n_{\rm th}$.  Different curves correspond to different values of $N$ in the $N$PE state.}
\label{NSvsN}
\end{figure}

Lastly we study the fraction of the total energy stored in the signal mode, i.e., $N_S/N$. 
As shown in Fig.~\ref{NSvsN}, the fraction of the total energy carried by the signal mode increases in proportion to $N$. That is, $N_S/N$ increases with $N$ for a given value of $n_{\rm th}$. %\CL{.}\rmv{, indicating that the fraction of the total energy carried by the signal mode increases with $N$.} % This reads the same as the previous sentence.
For $N \ge 30$ at least 90\% of the total energy is carried by the signal mode. 
   For $n_{\rm th} = 0$ the optimal state is always $|N,0\rangle$ so $N_S /N=1$, and the ratio $N_S/N$ quickly decreases with increasing $n_{\rm th}$ until $n_{\rm th} \approx 10$, from which point the changes become very slow. This indicates that it is beneficial to increase the fraction of the total energy carried by the idler mode as the noise level increases, but only up to a certain point. Past a modest noise level of $n_{\rm th} \approx 10$, the optimum fraction of the idler energy remains more or less the same, regardless of the total energy $N$.

For the coherent state, the optimal local measurement is known to be the homodyne measurement and in the asymptotic limit it coincides with the optimal global measurement for $n_{\rm th} \gg 1$ \cite{Guha}. The error probability in this case goes as $\exp[-(\eta^2 N_S M)/(4n_{\rm th}+2)]$, which is the same as the bound given by the QFI. Therefore, from the above results, we conclude that the performance of the $N$PE state is worse than that of the coherent state in the asymptotic limit, as long as local measurements are concerned. Furthermore, the optimal measurement generally depends on the probe state, so the measurement yielding $F_{\rm Q}^{N {\rm PE}}$ is likely to be complicated and dependent on $n_{\rm th}$.

Because the QFI for the 4PE state was higher than that of the coherent state in the scheme of Ref.~\cite{Lee21}, the authors considered a particular setup involving photon counting measurements. Under the non-optimal measurement scheme the SNR for an $N$PE state was shown to increase with $n_{\rm th}$. Such behaviour is surprising in light of the conclusion drawn above, so we investigate the measurement setup in detail in the next section.

\section{SNR for a non-optimal measurement scheme}
\begin{figure}[h]
\includegraphics[width=0.3\textwidth]{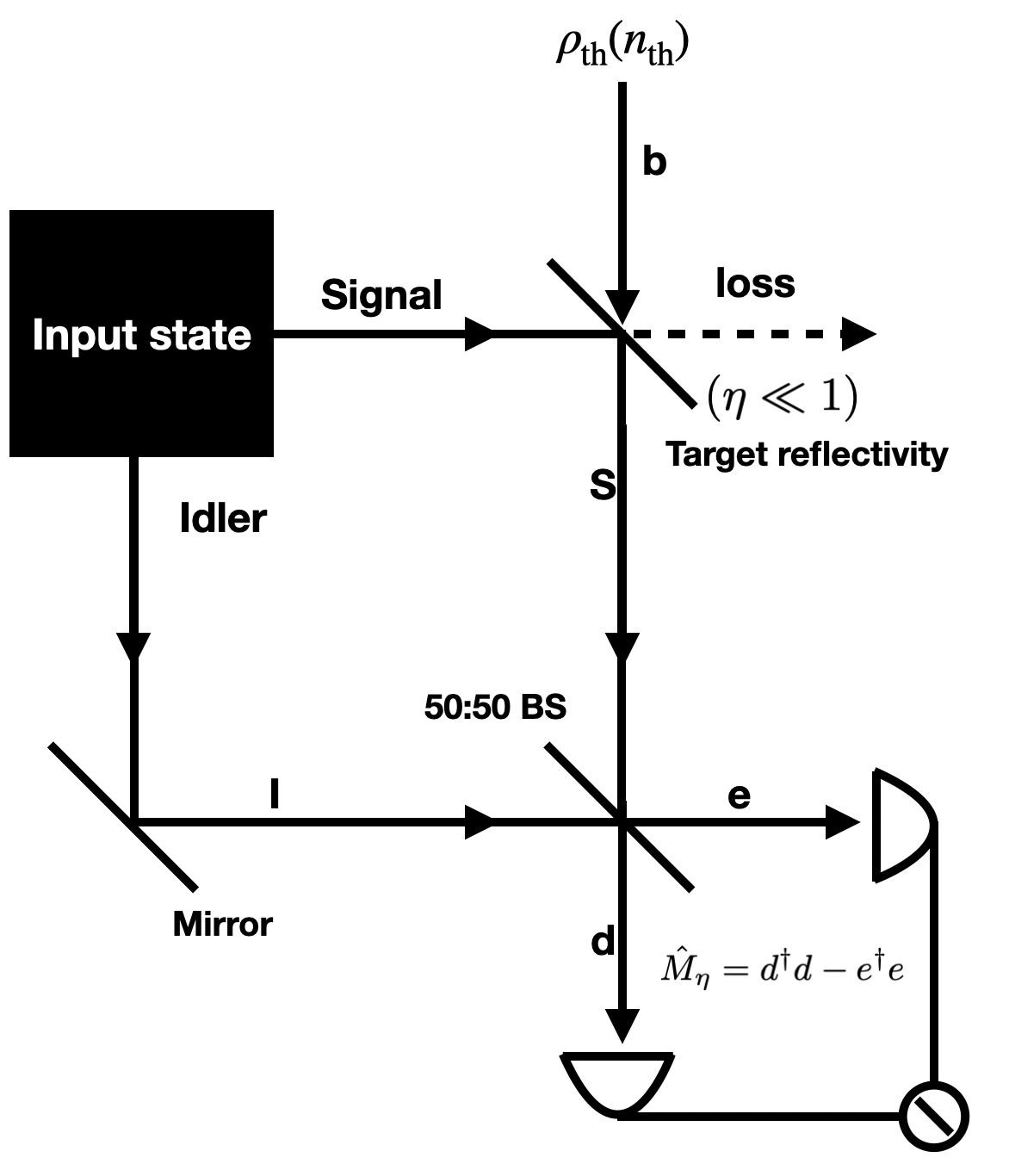}
\caption{Specific QI setup considered in Ref.~\cite{Lee21}. The input state is an optimized $N$PE state. For the measurement part, the reflected signal mode and the idler mode are mixed together by a 50:50 beam splitter and the number difference of the output modes is measured.}
\label{fig:setup_specific}
\end{figure}

Let us now consider the QI setup of Ref.~\cite{Lee21}, which is reproduced in Fig.~\ref{fig:setup_specific}. 
The measurement operator can be worked out straightforwardly and reads
\begin{align}
\hat{M}_\eta &= d^\dagger d - e^\dagger e 
\nonumber \\ &= \eta (a_I^\dagger a_S + a_I a_S^\dagger) + \sqrt{1-\eta^2}(a_I^\dagger b + a_I b^\dagger),
\end{align}
where $a_S$ and $a_I$ are the annihilation operators for the signal and idler modes, respectively.
For $\rho_\text{in} = |\psi\rangle_{SI}\langle\psi| \otimes \rho_{\rm th}$, we have
\begin{align}
\langle \hat{M}_\eta \rangle = \eta \langle a_I^\dagger a_S  + a_I a_S^\dagger\rangle_{SI}
\end{align}
and
\begin{align}
\langle \hat{M}^2 \rangle = \eta ^2 \langle ( a_I^\dagger a_S  + a_I a_S^\dagger)^2 \rangle + (1-\eta^2)  \langle (a_I^\dagger b + a_I b^\dagger)^2 \rangle.
\end{align}
The SNR for this measurement is
\begin{align}
{\rm SNR} = \frac{\langle \hat{M}_\eta \rangle - \langle \hat{M}_0\rangle }{\sqrt{(\Delta M_\eta)^2} +\sqrt{(\Delta M_0)^2}}.
\end{align}
Unlike the QFI,  the SNR depends on the idler energy as well as the signal energy, even for the separable coherent state input.  Therefore, for fair comparison, we fix the total energy $N$ and consider coherent states of the form $|\sqrt{N_S} e^{i\theta_S}, \sqrt{N-N_S}e^{i\theta_I}\rangle$.  %with the $N$PE states}. 
For such states one readily calculates the numerator to be $2\eta \sqrt{N_S(N-N_S)}  \cos\theta$, where $\theta \equiv \theta_S - \theta_I$, and in the denominator
\begin{align*}
(\Delta M_\eta)^2 =&   N-(1-\eta^2) N_S\\
&+ (2(N-N_S)+1)(1-\eta^2)n_{\rm th}. 
\end{align*}
The maximum value of the SNR is obviously at $\theta = 0$, so we set $\theta_S = \theta_I = 0$ without loss of generality. The optimization is performed over a real variable $N_S$.

The SNR for the 4PE state was claimed to be several times larger than that of the coherent state for  $n_{\rm th} \gtrsim 10$ in  Ref.~\cite{Lee21}. It was shown to decrease initially for small values of $n_{\rm th}$, but then to eventually increase linearly with $n_{\rm th}$. However we have seen that the maximum value of the SNR is proportional to the QFI, which  decreases with increasing $n_{\rm th}$. This makes the claimed behaviour puzzling. The apparent conflict results from the assumption of the separability of the thermal noise after the target reflection. Our exact calculation shows that the SNR decreases monotonically for an $N$PE state and is equal to that of the coherent state with the same signal energy.

\begin{figure}[t] 
\includegraphics[width=0.95\columnwidth]{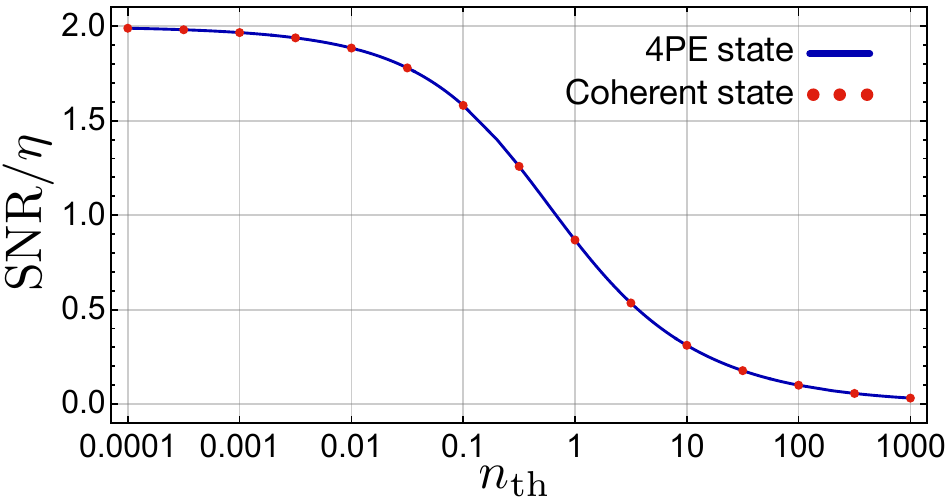}
\caption{Signal-to-noise ratio for the optimized 4PE state (solid blue curve) and  the corresponding result for coherent states with the same signal energy (red dots).}
\label{fig:SNR1}
\end{figure}

Figure \ref{fig:SNR1} plots the SNRs for the 4PE state and the coherent state when their signal energies are the same. The 4PE state has been optimized to maximize the SNR and the corresponding signal photon number was used to set $N_S$ for the coherent state. We note that the optimized coefficients are different to those maximizing the QFI, and that the optimization can be performed over real coefficients as shown in the Appendix.  Interestingly, two states yield indistinguishable SNRs for all values of $n_{\rm th}$ (the differences are in the order of $10^{-10}\sim10^{-5}$ and these results persist for larger values of $N$ (not shown here). Therefore, we conclude that in contrast to the findings in Ref.~\cite{Lee21}, $N$PE states show no perceptible advantage over the classical case even under the particular non-optimal measurement scheme.

\section{Discussion}
We have derived an achievable upper bound  and a loose lower bound on the minimum error probability, in the asymptotic limit and under repetitive local strategies. Both bounds are functions of the QFI, which are easier to calculate than the usual bounds in quantum hypothesis testing, such as the  quantum Chernoff bound.
As an application, we have considered definite-photon-number entangled states, which are known to be useful for quantum estimation purposes. We find that coherent states outperform the $N$PE states in the asymptotic limit, as far as local strategies are used.
%This is consistent with the known results, where the advantage holds for $N_S \ll 1$ and $n_{\rm th} \gg 1$, because in our case the signal energy is proportional to $N\gtrapprox 1$.  

We note that because the QFI is independent of an arbitrary unitary transformation on the idler mode, $N$PE states are equivalent to a general class of states in the Schmidt basis, which includes the truncated TMSV state. This result may seem to indicate that the infinite dimensional states are better than finite dimensional states for QI purposes. One must remember however, that $N$PE states have been optimized over all possible signal energies and that the TMSV states only outperform coherent states when $N_S \ll 1$.

In this regard, it is interesting to ask if $N$PE states can exhibit quantum advantage if $N_S$ is smaller than the optimal value that gives the maximum QFI. Preliminary results indicate that this is so. There are signal energy regimes in which $N$PE states perform better than the coherent states.  Detailed investigation on this will be reported in a future work.

\begin{acknowledgments}
CN acknowledges support by the National Research Foundation of Korea (NRF) grant funded by the Korea government(MSIT) (NRF-2019R1G1A1097074). 
CL is supported by a KIAS Individual Grant (QP081101) via the Quantum Universe Center at Korea Institute for Advanced Study.
SYL was supported by a grant to Defense-Specialized Project funded by Defense Acquisition Program Administration and Agency for Defense Development.
\end{acknowledgments}

\appendix 

\section{Proof of the optimality of real coefficients}
Here we prove that the optimal $N$PE states with respect to the SNR has real coefficients in the limit $\eta \ll 1$. To start with, note that the noise part of the SNR consists of terms independent of $\eta$ and those proportional to $\eta^2$. The dominant term is independent of $\eta$ and thus independent of the signal photon state. It only depends on $n_{\rm th}$ and the idler photon number $n_I$, which do not depend on the phases of the coefficients $\{ a_n \}$. On the other hand, the signal part of the SNR does depend on the relative phase differences because it is proportional to ${\rm Re}[\langle a_I^\dagger a_S \rangle]$. For the $N$PE states $\sum_n c_n|N-n,n\rangle$, the latter consists of terms of the form $|c_{n}||c_{n+1}|\cos (\theta_{n+1}-\theta_n)$. Because they are all positive terms which are added together, the maximum signal is achieved when all phases are equal.

\end{document}